\documentclass[conference]{IEEEtran}

\usepackage{cite}
\usepackage{amsmath,amssymb,amsfonts}
\usepackage{graphicx}
\usepackage{booktabs}
\usepackage{multirow}
\usepackage{array}
\usepackage{tabularx}
\usepackage{textcomp}
\usepackage{url}
\usepackage{stfloats}

\newcolumntype{Y}{>{\raggedright\arraybackslash}X}
\newcommand{\noegress}{\texttt{no\_egress}}
\newcommand{\sysname}{ContractWarden}

\def\BibTeX{{\rm B\kern-.05em{\sc i\kern-.025em b}\kern-.08em
    T\kern-.1667em\lower.7ex\hbox{E}\kern-.125emX}}

\begin{document}

\title{ContractWarden: Kernel-Enforced Damage Boundaries for AI Agents via Human-Authorized Contracts}

\author{
    \IEEEauthorblockN{
        Dongxu Cui\textsuperscript{1,2},
        Zhichao Gu\textsuperscript{2},
        Ping Zheng\textsuperscript{2},
        Wenshuai Xi\textsuperscript{2},
        Simeng Han\textsuperscript{2}, and
        Yong Liao\textsuperscript{1,*}}
    \IEEEauthorblockA{
        \textsuperscript{1}\textit{School of Cyber Science and Technology, University of Science and Technology of China}, Hefei, China\\
        cdx@mail.ustc.edu.cn, yliao@ustc.edu.cn}
    \IEEEauthorblockA{
        \textsuperscript{2}\textit{China Greatwall Technology Group Co., Ltd.}, Shenzhen, China\\
        \{guzhichao, zhengping, xiwenshuai, hansimeng\}@greatwall.com.cn\\
        ORCID: 0009-0007-1008-5465 (D. Cui), 0000-0001-6403-0557 (Y. Liao)\\
        \textsuperscript{*}Corresponding author: Yong Liao (yliao@ustc.edu.cn)}
}

\maketitle

\begin{figure*}[!b]
\footnotesize
This work has been submitted to the IEEE for possible publication.
Copyright may be transferred without notice, after which this version
may no longer be accessible.
\end{figure*}

\begin{abstract}
Large language model agents can execute commands, create subprocesses, and directly access files and networks, allowing prompt injection or planning errors to become operating-system side effects. We present \sysname, a Linux reference monitor that enforces a human-authorized damage boundary without trusting the agent or its policy suggestions. A model may propose a tri-state asset contract---\texttt{allow}, \texttt{deny}, or \noegress---but a human makes the final choice. An execution gate binds the contract to a concrete task before untrusted code runs. An extended Berkeley Packet Filter (eBPF) Linux Security Modules (LSM) data plane then enforces file and network decisions and monotonically propagates \noegress{} through processes, regular files, pipes, FIFOs, and supported Unix-domain sockets. All 570 runs across 19 security tests satisfy predefined return-value and side-effect criteria. On three co-located file-I/O workloads, median overhead is 11.96--12.89\% in a Linux 6.15 virtual machine and 35.79--61.54\% on a Linux 6.15 physical platform, lower than the evaluated frozen ActPlane baseline. The results demonstrate deterministic kernel enforcement for declared assets, supported paths, and controlled object lifecycles.
\end{abstract}

\begin{IEEEkeywords}
AI agents, human authorization, damage boundary, eBPF LSM, information-flow control, runtime enforcement
\end{IEEEkeywords}

\section{Introduction}

Large language model (LLM) agents have evolved into systems that plan tasks, invoke tools, execute scripts, and interact with local files and networks \cite{surveyLLM2024}. Indirect prompt injection, malicious tool output, or planning errors can therefore produce concrete system effects \cite{greshake2023,ASB,owasp2025,darkllm2025}. Application-layer validation is insufficient when one accepted tool call expands into a shell, interpreter, native binaries, subprocesses, temporary files, and direct system calls.

Three gaps make continuous enforcement difficult. First, attaching policy after process creation leaves a race in which untrusted code can run before registration. Second, process-tree inheritance misses lateral transfer through files or inter-process communication (IPC). Third, checking only network connection establishment misses a connection created before its process reads sensitive data. A restricted agent can otherwise launder its security state through a carrier that appears unrestricted.

We ask how an operating system can continuously enforce an asset damage boundary authorized before an untrusted task begins. \sysname{} separates policy proposal, semantic authorization, and privileged installation. A model only proposes a tri-state contract over a declared asset set; a human confirms every action; and a privileged control plane installs a complete, validated contract before releasing the task. Human confirmation is a trusted authorization root, not a claim that humans always classify assets correctly.

\sysname{} binds policy to a concrete kernel task using an execution gate and process file descriptor (pidfd). Its eBPF LSM data plane attaches source and \noegress{} state to kernel task and inode objects, monotonically merges state across supported process, file, pipe, FIFO, and Unix-socket edges, and checks egress at socket creation, connection, and send time. We make three contributions:
\begin{itemize}
    \item a human-authorized, closed-set asset contract that reduces a task damage boundary to bounded kernel actions;
    \item gate-before-exec binding and object-level monotonic propagation that resist state laundering; and
    \item an evaluation of 19 security properties, a four-configuration cost decomposition, and a co-located comparison with frozen ActPlane on virtual and physical platforms.
\end{itemize}

Our claims are scoped to declared assets, implemented kernel paths, and the object-lifecycle constraints described below.

\section{Background and Related Work}

Linux Security Modules mediate security-sensitive file, process, and socket operations; BPF LSM programs can deny an operation before it proceeds \cite{bpf-lsm-doc}. BPF maps retain enforcement state \cite{gregg2019bpf,bpf-maps-doc}, while task and inode local storage associate values with kernel objects rather than recycled numeric identifiers \cite{bpf-helpers-doc}. The verifier constrains program safety \cite{sun2024verifier}.

System provenance models processes and kernel objects as graph nodes and interactions as directed edges \cite{king2003backtracking,li2021survey}. CamFlow and CamQuery study whole-system capture and online analysis, while ProTracer, eAudit, and TAPAS reduce or structure audit volume \cite{camflow2017,camquery2018,Ma2016ProTracerTP,10646884,309640}. Dynamic information-flow systems instead propagate labels across computation and communication \cite{taintdroid2010}. AgentSight applies eBPF events to agent observability \cite{AgentSight2025}, and Agent-Warden uses object-lifetime state for process--file provenance in LLM-agent activity \cite{agent-warden}. These systems motivate object-level correlation, whereas \sysname{} targets authorization denial before a supported side effect.

AgentSpec, AgentBound, Progent, Fides, and CaMeL enforce policies in agent runtimes, server sandboxes, or trusted planner paths \cite{wang2026agentspec,Buhler2026AgentBound,progent2025,fides2025,camel2025}. They provide richer agent semantics and can complement a kernel monitor. Seccomp-BPF, AppArmor, and Landlock impose resource-oriented operating-system boundaries \cite{seccomp-doc,apparmor-doc,landlock-doc}, but do not automatically express the task-conditioned rule ``after reading this asset, propagate an egress restriction through an intermediate object.''

ActPlane is the closest system: it compiles project and task rules into operating-system policy with cross-event state and information-flow control \cite{actplane2026}. ActPlane targets programmable compliance policies; \sysname{} instead targets a small human-authorized asset boundary. We therefore compare overlapping properties without treating either system's non-overlapping semantics as failures.

\section{System Model and Design}
\label{sec:design}

\subsection{Contracts and Threat Model}

The system contains a human authorizer $H$, model proposer $M$, task domain $A_\tau$, privileged control plane $C$, and kernel enforcer $K$. The human maintains a declared asset set $\mathcal{R}$. For task $\tau$, the confirmed contract is the total function
\begin{equation}
\Gamma_\tau:\mathcal{R}\rightarrow
\{\texttt{allow},\texttt{deny},\texttt{no\_egress}\}.
\label{eq:contract}
\end{equation}
\texttt{allow} adds no \sysname{} restriction; \texttt{deny} rejects supported file and namespace operations; and \noegress{} permits local access but prevents a carrier that acquires the state from using managed network egress.

Each active task receives a non-reused user-space UUID and a bounded kernel position $\texttt{task\_id}\in[0,63]$, represented by source bit $b_\tau$. A process or supported object $x$ has state $S(x)=\langle L(x),E(x)\rangle$, where $L$ is a source bitmap and $E$ is the egress-denial bit. State joins monotonically:
\begin{equation}
\langle L_1,E_1\rangle\sqcup\langle L_2,E_2\rangle
=\langle L_1\lor L_2,E_1\lor E_2\rangle.
\label{eq:join}
\end{equation}

The proposer, task commands, agent inputs, and ordinary descendants are untrusted. They may fork, execute native code, rename files, pass descriptors, relay through supported IPC, connect before reading a restricted asset, and attempt identifier reuse. We trust the kernel, BPF/LSM path, privileged control plane, and human's final input. The task lacks root and kernel-management capabilities.

The design targets six security invariants. \emph{Closed authorization} requires every enabled asset to appear exactly once and prevents a model suggestion from directly becoming policy. \emph{Pre-execution binding} prevents untrusted code from running before installation and registration complete. \emph{Object identity} attaches state to kernel objects rather than only reusable path strings or process identifiers. \emph{Monotonic propagation} joins state on supported derivation, file, and IPC edges. \emph{Pre-side-effect denial} returns \texttt{EPERM} before a prohibited asset or network effect. Finally, \emph{fail-closed setup} keeps the child behind its gate if validation, installation, registration, or read-back fails.

\subsection{Lifecycle and Non-Goals}

When a root task exits, background descendants may remain. The launcher therefore tracks the derived process tree in a cgroup and cleans task policy only after the cgroup becomes empty. This condition terminates controlled processes but does not prove that a persistent inode or an out-of-cgroup receiver discarded an older source bit. A bounded source position cannot be safely reused until the implementation provides generation tags, complete global liveness accounting, or conservative non-reuse. Object-local storage does, however, make ordinary task and inode release independent of recycled numeric PIDs and inode numbers.

The monitor does not attempt to prove that an agent's plan is correct or that the human selected an optimal action. It does not protect undeclared assets, infer implicit or byte-level flows, mediate every IPC and device type, survive a compromised kernel, or provide an ordinary in-task declassification operation. These exclusions delimit the conditional invariant: given a complete confirmed contract, the supported side effects and propagation paths must obey it.

\subsection{Authorization and Pre-Execution Binding}

The model receives asset identifiers and metadata but need not receive canonical local paths. The control plane rejoins confirmed actions with its local asset table and rejects missing, duplicate, unknown, or out-of-domain actions. The proposer cannot authorize and the human does not directly mutate BPF maps.

The launcher creates a child blocked on a dedicated file descriptor, obtains a pidfd, and sends the pidfd, task identifiers, and contract to the daemon over an authenticated Unix \texttt{SOCK\_SEQPACKET} channel. The daemon validates peer credentials and pidfd/PID consistency, installs asset policy, attaches task state, and reads it back. Only then does the launcher open the execution gate. Failure leaves the workload blocked and triggers rollback. The pidfd identifies a concrete process object, while BPF task storage follows kernel-task lifetime.

\subsection{Compilation, Propagation, and Denial}

The control plane resolves an existing asset to a $(device,inode)$ key. Its value contains deny and \noegress{} source bitmaps; an update changes only one task's bit. Existing directory descendants are enumerated, while a bounded kernel ancestor walk covers later creations. Paths beyond 32 ancestors are outside the current scope.

Multiple tasks can safely share the long-lived data plane: installation for one task preserves every other active source bit on the same object. Directory compilation does not follow symbolic links into another tree and rejects a contract that expands past 4,096 policy keys. A local canonical path is therefore control-plane metadata; kernel decisions use object identity and a bounded ancestor relation.

For processes $P,Q$ and supported object $O$, the kernel performs
\begin{align}
S(Q)&\leftarrow S(Q)\sqcup S(P) && \text{on process derivation},\nonumber\\
S(O)&\leftarrow S(O)\sqcup S(P) && \text{on writable access},\nonumber\\
S(P)&\leftarrow S(P)\sqcup S(O) && \text{on readable access}.
\label{eq:propagation}
\end{align}
Regular files, FIFOs, and anonymous pipes use inode state. Because Unix-stream peers have different inodes, the sender joins state into a peer-state container and the receiver absorbs it. The permission-level approximation can over-propagate, but prevents ordinary code from clearing state by re-executing or relaying through an intermediate carrier.

A supported file operation is denied when the task's sources intersect the target inode's deny bitmap. The implementation covers open, create, truncate, unlink, and both rename endpoints. A \noegress{} match allows the file access while setting $E(P)=1$ in the same decision path. Egress checks at socket creation, connect, and send cover new sockets, connected TCP/UDP, established connections, and connectionless datagrams. Unix sockets are local propagation channels.

An \texttt{exec} changes the program image without clearing task state. Removing an asset rule stops future matches but cannot revoke \noegress{} already held by a live task or object. This monotonicity prevents an ordinary task from declassifying itself, although it can create false positives. A trusted release mechanism is deliberately left outside the task's authority.

Table~\ref{tab:impl} summarizes the state and enforcement path. A cross-process lock serializes writers; the loader validates schema, asset type, bit consistency, action disjointness, and map layouts, then snapshots and restores affected keys on update failure. Audit events use a ring buffer independent of the LSM return value, so audit pressure cannot turn deny into allow.

\begin{table}[t]
\caption{Prototype State, Hooks, and Bounds}
\label{tab:impl}
\centering
\scriptsize
\setlength{\tabcolsep}{3pt}
\begin{tabularx}{\columnwidth}{@{}p{0.22\columnwidth}Yp{0.25\columnwidth}@{}}
\toprule
\textbf{State/path} & \textbf{Purpose} & \textbf{Bound/hooks} \\
\midrule
Task storage & Source and egress state & 64 source positions \\
Inode storage & File and IPC propagation & Object lifetime \\
Asset policy & Deny/egress bitmaps & 4,096 keys; 32 ancestors \\
Unix-peer map & Peer receive state & 4,096 keys \\
File LSM & Access and propagation & open, permission, create, truncate, unlink, rename \\
Socket LSM & Egress and Unix flow & create, connect, send, receive \\
Process trace & Inherit and audit & fork and exit \\
Event ring & Asynchronous evidence & 256 KiB plus drop count \\
\bottomrule
\end{tabularx}
\end{table}

\begin{table*}[t]
\caption{eBPF Attachment Points and Responsibilities}
\label{tab:hooks}
\centering
\scriptsize
\begin{tabularx}{0.98\textwidth}{@{}p{0.22\textwidth}Yp{0.22\textwidth}Y@{}}
\toprule
\textbf{Attachment} & \textbf{Responsibility} & \textbf{Attachment} & \textbf{Responsibility} \\
\midrule
\texttt{lsm/file\_open} & Asset decision and regular-file propagation & \texttt{lsm/file\_permission} & Pipe/FIFO flow and socket-fd write check \\
\texttt{lsm/inode\_create} & Protected directory creation denial & \texttt{lsm/path\_truncate} & Truncation denial \\
\texttt{lsm/inode\_unlink} & Unlink denial & \texttt{lsm/inode\_rename} & Source and destination rename checks \\
\texttt{lsm/socket\_create} & New external-socket denial & \texttt{lsm/socket\_connect} & Destination connection denial \\
\texttt{lsm/socket\_sendmsg} & Established-send denial and Unix send flow & \texttt{lsm/socket\_recvmsg} & Unix receiver absorption \\
\texttt{sched\_process\_fork} & Child-state inheritance & \texttt{sched\_process\_exit} & Exit audit \\
\texttt{security\_inode\_free} & Hash-key cleanup at inode release & & \\
\bottomrule
\end{tabularx}
\end{table*}

\section{Implementation and Evaluation}

\subsection{Prototype and Method}

The prototype comprises ordinary-user asset and contract tools, a C launcher and privileged daemon, a policy loader, and a CO-RE BPF data plane. It requires BPF LSM, BPF Type Format, bpffs, cgroup v2, pidfd, and task/inode local storage. The daemon maintains one BPF instance for multiple tasks. Task storage is attached to the thread-group leader so threads share logical state. The loader uses bounded JSON parsing and verifies contract version, task bit, asset identity and type, action-set disjointness, and BPF map key/value layouts. Table~\ref{tab:hooks} lists the enforcement path.

We separately evaluate security behavior, mechanism cost, and a frozen ActPlane baseline. Model calls and human interaction are outside timing; performance runs use deterministic proposals but traverse contract validation, persistence, policy loading, gating, and registration. \sysname{} security and cost decomposition run on Linux 7.0. The baseline property tests use a Linux 6.15 VMware guest because frozen ActPlane requires that kernel. Co-located performance runs under Linux 6.15 both in the eight-vCPU guest and directly on an AMD Ryzen AI MAX+ 395 platform with 32 logical processors. We compare absolute times only within a platform batch.

Each performance configuration has five warmups and 30 measured repetitions in randomized within-repetition order. For system $s$ and repetition $r$, paired overhead is
\begin{equation}
O_{s,r}=\frac{T_{s,r}-T_{N,r}}{T_{N,r}}\times100\%,
\end{equation}
where $T_{N,r}$ is Native time in the same repetition. Intervals use 10,000 bootstrap resamples of the 30 paired overheads.

\begin{table*}[t]
\caption{Co-Located File-I/O Performance (Median of 30 Paired Trials)}
\label{tab:perf}
\centering
\scriptsize
\setlength{\tabcolsep}{3.5pt}
\begin{tabular}{@{}llrrrrr@{}}
\toprule
\textbf{Platform} & \textbf{Workload} & \textbf{Native ms} & \textbf{ActPlane ms} & \textbf{AP overhead [95\% CI]} & \textbf{\sysname{} ms} & \textbf{CW overhead [95\% CI]} \\
\midrule
\multirow{3}{*}{VM} & Public read & 14864.1 & 19807.8 & 33.39\% [33.04,33.81] & 16617.4 & 11.96\% [11.64,12.14] \\
 & Protected read & 14948.1 & 24051.6 & 61.02\% [60.70,61.11] & 16862.4 & 12.89\% [12.66,13.05] \\
 & 14 assets & 10875.7 & 16612.5 & 52.44\% [52.13,53.16] & 12210.9 & 12.32\% [11.90,12.64] \\
\midrule
\multirow{3}{*}{Physical} & Public read & 810.1 & 1916.4 & 136.94\% [135.89,138.24] & 1281.7 & 58.32\% [57.74,58.91] \\
 & Protected read & 830.4 & 2245.1 & 170.72\% [169.78,172.18] & 1338.0 & 61.54\% [60.49,62.37] \\
 & 14 assets & 994.0 & 2775.0 & 179.03\% [178.25,181.35] & 1351.2 & 35.79\% [35.48,36.38] \\
\bottomrule
\end{tabular}
\end{table*}

\begin{figure*}[t]
\centering
\includegraphics[width=0.90\textwidth]{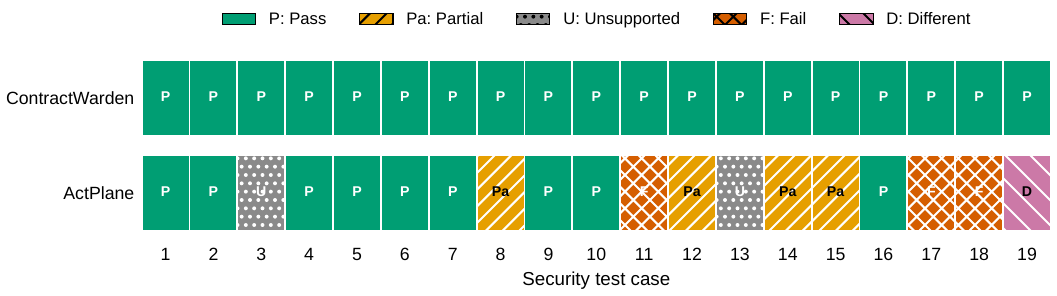}
\caption{Outcomes for the 19 target properties. \sysname{} results use 30 repetitions per property; frozen ActPlane uses one execution per property.}
\label{fig:security-matrix}
\end{figure*}

\begin{figure}[t]
\centering
\includegraphics[width=\columnwidth]{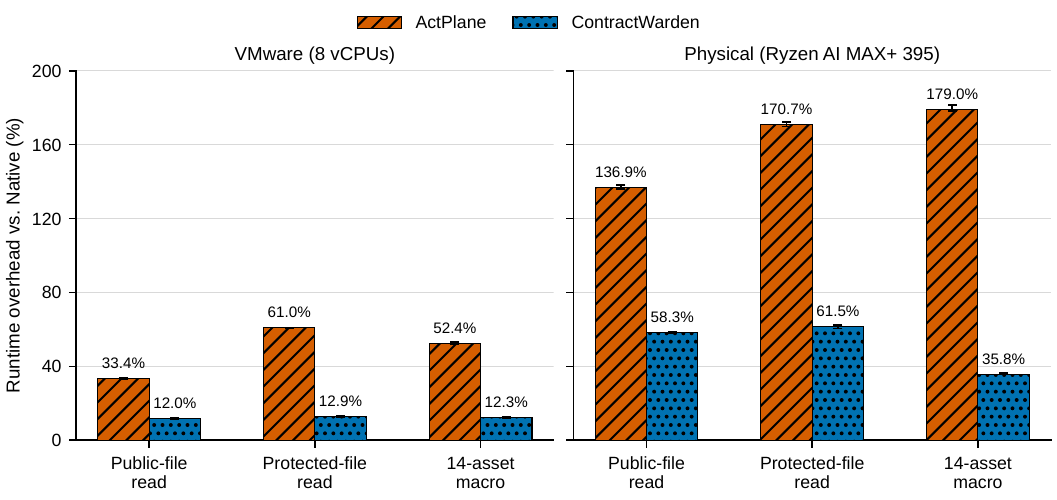}
\caption{Median paired runtime overhead relative to Native, with 95\% bootstrap intervals.}
\label{fig:performance}
\end{figure}

Figures~\ref{fig:security-matrix} and~\ref{fig:performance} summarize the property outcomes and paired overheads that the following subsections analyze.

\subsection{Security Correctness}

Security tests use system-call return values and resource side effects as primary criteria, rather than audit events alone. Each repetition randomizes all 19 cases and cleans controlled processes and policy between cases. All 570 observations (19 cases $\times$ 30 repetitions) pass, and the aggregate ring-buffer drop count is zero. Table~\ref{tab:security} groups the tested properties.

\begin{table*}[t]
\caption{Security Properties and Results}
\label{tab:security}
\centering
\scriptsize
\setlength{\tabcolsep}{3pt}
\begin{tabularx}{0.98\textwidth}{@{}r p{0.24\textwidth}Y r@{}}
\toprule
\textbf{No.} & \textbf{Property} & \textbf{Primary behavioral criterion} & \textbf{Pass/Run} \\
\midrule
1 & Source-isolated \texttt{deny} & Restricted task gets \texttt{EPERM}; other source reads & 30/30 \\
2 & Public asset & Network succeeds without restricted state & 30/30 \\
3 & Post-infection established send & Send succeeds before and fails after infection & 30/30 \\
4 & Sibling isolation & Infected branch denied; sibling remains usable & 30/30 \\
5 & Fork inheritance & Child inherits \noegress & 30/30 \\
6 & Cross-task regular file & A writes, B reads, then B cannot network & 30/30 \\
7 & No reverse flow on exit & Restricted child exit does not infect parent & 30/30 \\
8 & Hot update and monotonicity & New rule applies; removal retains acquired state & 30/30 \\
9 & FIFO propagation & A--FIFO--B causes B egress denial & 30/30 \\
10 & 64-bit source isolation & 32 targets denied; 32 controls allowed & 30/30 \\
11 & File-operation coverage & Write/create/truncate/rename/unlink leave no forbidden effect & 30/30 \\
12 & Aliases and inode lifetime & Links/rename do not bypass; reused inode is clean & 30/30 \\
13 & Established-socket sends & \texttt{send}, \texttt{sendto}, \texttt{sendmsg}, write denied & 30/30 \\
14 & Invalid setup and rollback & Six invalid/injected failures do not start target & 30/30 \\
15 & PID reuse & New process does not inherit old task state & 30/30 \\
16 & Anonymous pipe & Producer--pipe--consumer propagates restriction & 30/30 \\
17 & Unix stream peer & Receiver absorbs sender state & 30/30 \\
18 & Socketpair peer & Receiver absorbs sender state & 30/30 \\
19 & Directory ancestor bound & Depth 31 matches; depth 32 is outside scope & 30/30 \\
\bottomrule
\end{tabularx}
\end{table*}

Cases 3 and 13 establish the need for send-stage enforcement. Case 11 verifies both return values and filesystem effects; case 14 covers pre-execution rejection and rollback. Thus, within the implemented paths, the monitor preserves source isolation, monotonic propagation, and pre-side-effect denial.

The cases exercise four complementary boundaries. Isolation tests separate sources, siblings, and reused identities. Propagation tests cover process derivation and every currently supported persistent or transient carrier. Enforcement tests check both newly created and already established network channels, including all four send interfaces exercised by the harness. Lifecycle tests cover rule replacement, failed installation, object release, PID reuse, and the documented directory-depth boundary. This partition matters because a pass on simple fork inheritance does not imply that file-mediated or peer-socket laundering is prevented.

\subsection{Cost and Frozen-Baseline Comparison}

The cost decomposition uses Baseline (not loaded), Loaded-idle (BPF loaded, measured task unregistered), Tracing (task registered with allow-all policy), and Full (asset policy enabled). Two microbenchmarks execute one million \texttt{open/read/close} transactions; a 14-asset macrobenchmark executes 700,000. All 420 trials are valid. Median Full overhead is 21.16--35.62\%; most of the observed increase is between Loaded-idle and Tracing, while the Tracing--Full difference is smaller. We do not interpret this observation as zero policy-lookup cost.

\begin{table}[t]
\caption{File-I/O Cost Decomposition}
\label{tab:cost}
\centering
\scriptsize
\setlength{\tabcolsep}{3pt}
\begin{tabular}{@{}llrr@{}}
\toprule
\textbf{Workload} & \textbf{Configuration} & \textbf{Median ms} & \textbf{Paired \%} \\
\midrule
\multirow{4}{*}{Public $10^6$} & Baseline & 764.663 & -- \\
 & Loaded-idle & 871.964 & 14.077 \\
 & Tracing & 1019.968 & 33.592 \\
 & Full & 1019.965 & 33.230 \\
\midrule
\multirow{4}{*}{Protected $10^6$} & Baseline & 787.789 & -- \\
 & Loaded-idle & 889.757 & 13.220 \\
 & Tracing & 1058.295 & 34.788 \\
 & Full & 1069.548 & 35.622 \\
\midrule
\multirow{4}{*}{14 assets} & Baseline & 962.830 & -- \\
 & Loaded-idle & 1034.648 & 7.666 \\
 & Tracing & 1156.708 & 20.287 \\
 & Full & 1165.788 & 21.156 \\
\bottomrule
\end{tabular}
\end{table}

Frozen ActPlane first passes a preflight confirming real eBPF loading and \texttt{EPERM} denial on Linux 6.15. Against the same 19 target properties, one execution per property yields 9 Pass, 3 Fail, 4 Partial, 2 Unsupported, and 1 Different. Non-pass outcomes primarily reflect a missing send-stage enforcement point, creation leaving an empty file before denial, absent Unix-peer propagation, and different update, identity, or directory semantics. These categories indicate partial comparability or different expression points, not ten generic failures; the properties also do not cover all ActPlane temporal-policy features.

For co-located performance, Native, ActPlane, and \sysname{} use identical benchmark code, files, counts, and content hashes. Table~\ref{tab:perf} reports medians from 30 measured trials per cell. All 630 trials (including warmups) are valid; hashes match, and the maximum drop count across 210 \sysname{} daemon lifecycles is zero.

\sysname{} overhead is 11.96--12.89\% in the eight-vCPU VMware guest and 35.79--61.54\% on the 32-thread AMD Ryzen AI MAX+ 395 platform, versus 33.39--61.02\% and 136.94--179.03\% for frozen ActPlane. Bootstrap intervals do not overlap within either platform. The higher percentages on the faster physical baseline show why platforms must be interpreted separately; the result is not an overall feature ranking.

The virtual and physical batches agree on the ordering but not on percentage magnitude. The same fixed per-operation work occupies a larger fraction of a faster native runtime, and the guest additionally includes hypervisor scheduling. Accordingly, we draw only within-batch conclusions. Timing covers steady-state workload execution and excludes engine startup, model proposal, and human confirmation; the latter are deployment and usability costs rather than per-operation data-plane costs.

\section{Discussion and Limitations}

\subsection{Authorization and Declassification}

The guarantee is conditional on the human-confirmed contract: an omitted asset or permissive choice remains possible. Human confirmation determines who authorizes the boundary, not whether the judgment is objectively correct. User studies are needed for usability, decision time, alert fatigue, and contract-quality claims, but are distinct from testing whether a fixed contract becomes kernel denial.

Permission-level propagation may approximate the ability to write as actual influence and therefore over-taint a task. The trade-off keeps state bounded while reducing missed propagation through intermediate carriers. Restoring egress should require a trusted declassification mechanism, such as an attested transformation or independently authorized release; ordinary task code must not remove a label merely by declaring data safe.

\subsection{Lifecycle and Policy Publication}

The 64-bit source representation makes joins fixed-width but bounds the active namespace. Cgroup drainage proves that the derived task tree exited; it does not prove every persistent inode or out-of-cgroup receiver discarded a source bit. Safe reuse requires generation tags, global liveness tracking, or conservative non-reuse. Multi-key updates are serialized and rolled back on failure but are not linearizable to concurrent kernel readers; double-buffered maps with a one-key generation switch are a natural extension.

\subsection{Scope and External Validity}

The prototype does not cover undeclared assets, implicit or byte-level flows, every IPC/device class, a root-compromised host, general in-task declassification, or destination allowlists. A deployment can compose a container or Landlock for maximum privilege, \sysname{} for task-conditioned assets, and an agent runtime for semantic feedback. One physical machine, one VM, and file-I/O microbenchmarks do not represent all agent workloads. Random order, warmups, content hashes, and pairing reduce temporal noise but do not eliminate VM scheduling effects. Network, fork/IPC, update, concurrent-task, multi-kernel, and realistic agent workloads remain evaluation extensions.

The 19 properties derive from our damage-boundary threat model and are therefore better aligned with \sysname{} than with the full ActPlane policy language. The frozen comparison uses only one execution per ActPlane property and supports a coverage observation, not a repeated pass-rate estimate. Security, decomposition, and co-located measurements also come from separate kernel batches; we never compare absolute time across those batches. The physical run removes hypervisor scheduling as a confounder but still represents one machine. These design choices bound the evidence without changing the within-batch paired measurements or the repeated \sysname{} security criteria.

\section{Conclusion}

Within this scope, \sysname{} turns a human-authorized asset contract into task-bound kernel state. Gate-before-exec registration prevents a startup race; object-level monotonic propagation prevents supported carrier laundering; and file plus multi-stage network hooks deny prohibited side effects. All 570 security observations pass, while co-located tests show lower overhead than the evaluated frozen ActPlane configuration. Future work will add generation-safe source reuse, atomic policy publication, trusted declassification, and realistic concurrent agent workloads.

\section*{AI Disclosure}

During preparation, the authors used Google Gemini, and OpenAI Codex for English editing, manuscript organization, and readability. The authors reviewed all AI-assisted content and take responsibility for the manuscript, citations, and results.

\bibliographystyle{IEEEtran}
\bibliography{ContractWarden}

\end{document}